# NaCo$_2$(SeO$_3$)$_2$(OH): Competing Magnetic Ground States of a New Sawtooth Structure with 3$d^7$ Co$^{2+}$ Ions

Liurukara D. Sanjeewa,[a,b]* V. Ovidiu Garlea,[c] Keith M. Taddei,[c] Li Yin,[d] Jie Xing,[d] Randy S. Fishman,[d] David S. Parker,[d] Athena S. Sefat[d]

[a]University of Missouri Research Reactor (MURR), Columbia, Missouri 65211, USA
[b]Department of Chemistry, University of Missouri, Columbia, Missouri 65211, USA
[c]Neutron Scattering Division, Oak Ridge National Laboratory, Oak Ridge, TN 37831, USA
[d]Materials Science and Technology Division, Oak Ridge National Laboratory, Oak Ridge, TN 37831, USA

*Corresponding Authors

## Abstract

While certain magnetic sublattices have long been known theoretically to give rise to emergent physics via competing magnetic interactions and quantum effects, finding such configurations in real materials is often deeply challenging. Here we report the synthesis and characterization of a new such material, NaCo$_2$(SeO$_3$)$_2$(OH) which crystallizes with a highly frustrated sublattice of sawtooth Co$^{2+}$ chains. Single crystals of NaCo$_2$(SeO$_3$)$_2$(OH) were synthesized using a low-temperature hydrothermal method. X-ray single crystal structure analysis reveals that the material crystallizes in orthorhombic space group of *Pnma* (no. 62). Its crystal structure exhibits one-dimensional chains of corner-sharing isosceles triangles that are made of two crystallographically distinct 3$d^7$ Co$^{2+}$ sites (Co(1) and Co(2)). The chains run along the *b*-axis and are interconnected via [SeO$_3$] groups to form a three-dimensional structure mediating super-exchange interactions. The temperature dependent magnetization data show a ferromagnetic-like (FM) transition at 11 K ($T_1$) followed by an antiferromagnetic (AFM) transition at about 6 K ($T_2$). Neutron-powder diffraction measurements reveal that at $T_1$ = 11 K only Co(2) site orders magnetically, forming ferromagnetic zigzag chains along the *b*-axis. Below $T_2$ = 6 K, both Co(1) and Co(2) sites order in an nearly orthogonal configuration, with Co(1) moments lying inside the plane of the sawtooth chain while Co(2) moments cant out of the plane. The canting of the magnetic moments leads to a net ferromagnetic component along *b*-axis, parallel to the chain direction. The ordered moments are fully compensated in the *ac*-plane. Inelastic neutron scattering measurements reveal crystal field excitations that are consistent with the presence of a spin-orbital entangled pseudo-spin state $J_{eff} = 1/2$ for the $Co^{2+}$ ions. Low-energy spin-wave excitations are observed below the second magnetic transition. The analysis of powder excitation spectrum suggests complex exchange interactions that go beyond a Heisenberg Hamiltonian model with nearest neighbor couplings. Our results demonstrate the richness of the magnetic properties of sawtooth-type structures and encourage the study of similar structures with different oxyanion groups.

1. Introduction

The search for new classes of quantum materials is a vigorous pursuit among material scientists and condensed matter physicists due to the powerful implications many such materials have for technological applications. Specifically, due to their potential to host exotic states of matter with novel transport properties robust to perturbations, such as the quantum spin-liquid state, quantum materials have attracted special attention in the past decade due to the interest of quantum information applications. QSLs exhibit a highly entangled quantum state which can arise in frustrated magnetic systems, typically with specific magnetic sublattices (e.g. Kagome, triangular, honeycomb) which give rise to unique bond directional interactions and inherent geometric frustration. Such configurations can lead to numerous degenerate magnetic ground states which prevent the formation of long-range magnetic order and invoke topological classifications with exotic excitations exhibiting non-abelian statistics. Examples of such systems have focused mainly in two-dimensional (2D) triangular, Kagome and honeycomb magnetic lattices and three-dimensional (3D) pyrochlore lattices. The triangular magnetic lattice, in which magnetic ions reside on the vertices of the triangles, is one of the simplest ways to achieve geometric frustration. In this geometry long-range magnetic order can be suppressed by an inability of the system to satisfy all magnetic interactions simultaneously. From the theoretical point of view, such frustration could manifest quantum fluctuations and lead to a spin-liquid like state near $T = 0$ K.[1-8]

In addition to the 2D triangular magnetic lattices, geometrical frustration has also been observed in one-dimensional (1D) chains where the topology of corner sharing antiferromagnetic triangles form infinite chains. For example, the triangular spin tube, sawtooth (Δ-chain) chain and half sawtooth chains are seminal models in the field of frustrated 1D quantum magnetic systems.[9-12] In the sawtooth chain magnetic lattice, the corner-sharing magnetic triangles connect in an up and down fashion along the chain. Within each triangle, two crystallographically independent sites produce two different exchange parameters between the base-base ($J_{bb}$) and the base-vertex ($J_{bv}$) pathways. Therefore, complex magnetic ground states can be anticipated depending upon the relative strengths of $J_{bb}$ and $J_{bv}$ such as Ising-type interactions, Heisenberg interactions, and Dzyaloshiniskii-Moriya interactions leading to rich magnetic phase diagrams. The ground state of a uniform $S = 1/2$ sawtooth chain is theoretically well understood. It consists of a twofold degenerate superposition of spin singlets formed by either left or right pair of spins of each triangle. The lowest excitation in a sawtooth chain with periodic boundary conditions is given by a *kink-antikink* pair, which has a dispersionless gap $\Delta E = 0.234J$ where $J$ is the coupling between pairs of spins.[9-20] Moreover, in the classical limit, some saw-tooth model systems possess zero energy flat-band modes, remarkably similar to a Kagome antiferromagnet (sawtooth magnetic lattice is a partial structure of the Kagome lattice). Strongly correlated flat-band systems with localized magnon states in high magnetic fields are the subject of intensive studies, as flat-band modes offer promising possibilities for manipulating the propagation of waves of any origin. Therefore, though originally driven by the search for complex magnetic ground states, recent studies on sawtooth chain lattices have been devoted to developing magnonic devices due to the presence of highly degenerate flat bands.[21]

To date, the realization of sawtooth lattice structures has been very limited to: delafossite-$YCuO_{2.5}$,[22-23] olivines-$ZnLn_2S_4$ ($Ln$ = Er and Yb),[24] $A_2BX_4$ ($A$ = Mn, Fe, Ni; $B$ = Si, Ge; $X$ = S, Se, Te O),[25-29] euchroite-$Cu_2(AsO_4)(OH)$ $3H_2O$,[30-31] $Rb_2Fe_2O(AsO_4)_2$,[32] $Cu_2Cl(OH)_3$,[14] and $Fe_2O(SeO_3)_2$[33]. Experimentally, $YCuO_{2.5}$,[22] $Cu_2(AsO_4)(OH)$ $3H_2O$[29] and $ZnLn_2S_4$[23] do not show any long range ordering, and are therefore thought promising candidates to study quantum spin liquid behavior. On the other hand, oxyanion-based compounds, $Rb_2Fe_2O(AsO_4)_2$ ($T_N$ = 25 K)[32] and $Fe_2O(SeO_3)_2$ display interesting magnetic properties, with the former having a complex magnetic phase diagram with field induced states and the latter exhibiting an unusually high long-range magnetic ordering temperature for this type of compound, $T_C$ = 105 K[33]. The $A_2BX_4$ ($A$ = Mn, Fe, Ni; $B$ = Si, Ge; $X$ = S, Se, Te, O) olivine type structures display a wide range of magnetic ground states depending upon the $A$-, $B$- and $X$-sites ions[25-29]. Hence, the search for novel magnetic grounds states has made the synthesis of new sawtooth compounds a goal for materials chemists. The synthesis of transition metal-based oxyanions compounds is mostly limited to tetrahedral groups such as $[PO_4]^{3-}$, $[AsO_4]^{3-}$, $[VO_4]^{3-}$, $[MoO_4]^{2-}$. These oxyanion groups together with open shell transition metal ions can form a variety of low-dimensional structures. In general, single crystals of these transition metal-based oxyanions are synthesized using fluxes (salt fluxes, carbonate fluxes, oxide fluxes and etc.).[34-37] Recently, we have been exploring novel transition metal-based oxyanions quantum materials using both low-temperature ($T$ < 220 °C) and high-temperature ($T$ = 500 – 700 °C) hydrothermal synthesis.[38] We have successfully employed the hydrothermal method to synthesize numerous first-row transition metal $[VO_4]^{3-}$ and $[MoO_4]^{2-}$ compounds which encouraged us to explore transition metal-based selenium(IV) oxyanions compounds and perform detailed physical property studies.[39-51] The realization of selenium(IV) oxyanions compounds targeting exotic magnetic properties has been very limited and only a handful of reports are available in the literature. Furthermore, selenium can adapt $[SeO_3]^{2-}$ and $[Se_2O_5]^-$ oxyanion configuration which can form a great number of different structural arrangements in extended solid structures providing a rich and diverse crystal chemistry. Therefore, we now extend our hydrothermal synthesis to scrutinize $[SeO_3]$-based transition metal compounds and perform detailed physical property characterizations and magnetic structure determination using neutron diffraction searching for novel magnetic ground states. To give one example, recently we performed comprehensive magnetic and neutron diffraction studies on $TM_3(SeO_3)_3$ $H_2O$ ($TM$ = Mn, Co, Ni) compounds which exhibited tunable magnetic ground states with metamagnetic transitions.[48]

In this paper, we report the hydrothermal synthesis of $NaCo_2(SeO_3)_2(OH)$ single crystals and a comprehensive characterization using both bulk probes and neutron scattering techniques. Hydrothermal synthesis of both $NaCo_2(SeO_3)_2(OH)$ (*Pbnm*) and $NaZn_2(SeO_3)_2(OH)$ (*Pnma*) were reported previously.[52-53] The magnetic phase diagram of the system was studied using multiple experimental methods including magnetic susceptibility, isothermal magnetization and specific heat. Magnetic susceptibility and heat capacity measurements reveal the presence of two competing magnetic transitions which is very unusual for this type of sawtooth oxyanion compounds while isothermal magnetization displays a field induced metamagnetic transition. Temperature dependent neutron powder diffraction was used to determine the magnetic states associated with each of the two magnetic transitions showing that below the first transition only some of the Co sites order with the second transition leading to ordered magnetic moments on both Co sites. Inelastic neutron scattering measurements (Supporting Information) over a broad range of energy transfers reveal crystal field levels indicative of a spin-orbital entangled $J_{eff} = 1/2$ state. Measurements at low energy transfers show spin-wave excitations which only become well defined below the second transition with only a diffuse signal observed between the first and second transition temperatures consistent with the partial ordering found in the diffraction measurements. Together these results suggest a complex magnetic phase diagram with evidence of spin-orbital entangled $J_{eff} = 1/2$ state for $Co^{2+}$ in $NaCo_2(SeO_3)_2(OH)$.

## 2. Experimental Section

### 2.1 Hydrothermal Synthesis of $NaCo_2(SeO_3)_2(OH)$

$NaCo_2(SeO_3)_2(OH)$ was synthesized using a low-temperature hydrothermal method. Here, we used two different synthesis methods to grow sizable single crystals for single crystal X-ray diffraction and a bulk microcrystalline sample to perform the physical property characterization and neutron scattering experiments. In the first method, a total of 0.4 g of $NaHCO_3$, $Co(OH)_2$ and $H_2SeO_3$ were mixed in a stoichiometric ratio of 5 : 2: 2 with 8 mL of water and loaded into a Teflon-lined stainless-steel autoclave which was then well sealed. The reaction mixture was heated at 200 °C for 7 days. After the reaction, columnar crystals (0.2 mm, Figure SI1) were recovered using suction filtration by washing with de-ionized water and acetone. The yield of this method was very low and also grew other impurity phases. Therefore, a different approach was employed to grow a high-yield of micro-crystalline samples of $NaCo_2(SeO_3)_2(OH)$. In this method, a total of 1.2 g of $NaHCO_3$, $Co(OH)_2$ and $H_2SeO_3$ were mixed in a stoichiometric ratio of 1 : 2 : 2 with 12 mL of water and loaded into a Teflon-lined stainless-steel autoclave, sealed well and then heated at 200 °C for 14 days. After cooling to room-temperature, a purple color microcrystalline sample was recovered using filtration. Several sets of reactions were performed to obtained 5 g of powder sample to perform the neutron powder diffraction experiments. Purity of the samples were checked using the powder X-ray diffraction method.

## 2.2 Single Crystal X-ray Diffraction

Single crystals of $NaCo_2(SeO_3)_2(OH)$ were sonicated in water to remove any surface impurities before determining the crystal structure using single crystal X-ray diffraction (SXRD). The SXRD was performed by Bruker Quest D8 single-crystal X-ray diffractometer. The data were collected at room temperature utilizing a Mo Kα radiation, λ = 0.71073 Å. The crystal diffraction images were collected using $\phi$ and $\omega$-scans. The diffractometer was equipped with an Incoatec IμS source using the APEXIII software suite for data setup, collection, and processing.[55] The structure was resolved using intrinsic phasing and full-matrix least square methods with refinement on $F^2$. Structure refinements were performed using the SHELXTL software suite.[56] All atoms were first refined with isotropic displacement parameters which were then converted to anisotropic displacement parameters and allowed to refine. Additionally, energy-dispersive spectroscopy analysis (EDS) was performed using a Hitachi S-3400 scanning electron microscope equipped with an OXFORD EDX microprobe to confirm the elemental composition in single crystal samples.

## 2.3 Powder X-ray Diffraction

Room-temperature powder x-ray diffraction data were collected using a PANalytical X'Pert Pro MPD diffractometer with Cu Kα1 radiation (λ = 1.5418 Å). PXRD pattern of ground single crystals samples were collected to confirm the purity of the samples and Figure SI2 shows the powder patterns of $NaCo_2(SeO_3)_2(OH)$.

## 2.4 Magnetic Property Characterization

Temperature-dependent and field dependent magnetic susceptibility measurements were performed using a Quantum Design Magnetic Property Measurement System (MPMS). Here, a ground single crystal sample was pressed into a 1/8 inch pellet and affixed to a quartz rod using GE varnish. The temperature dependent magnetization measurements were carried out using 6.2 mg pellet from 2 to 350 K in an applied magnetic field of up to 50 kOe. Additionally, isothermal magnetization measurements were performed between 2-100 K up to a 60 kOe magnetic field. The heat capacity ($C_p$) of the sample was measured using a Physical Property Measurement System between 2-50 K under 0 and 110 kOe applied magnetic field.

## 2.5 Neutron Scattering

Neutron powder diffraction (NPD) measurements were performed on a $NaCo_2(SeO_3)_2(OH)$ polycrystalline sample, obtained from ground single crystals using the HB2A Powder Diffractometer at High Flux Isotope Reactor (HFIR) at Oak Ridge National Laboratory.[57] Diffraction patterns were collected using the open-21'-12' collimator settings (for pre-monochromator, pre-sample and pre-detector collimation, respectively) and with both 1.54 and 2.41 Å incident wavelength settings to optimize both broad $q$ coverage as well as access low scattering angles. A powder sample with a total mass of approximately 5 g was loaded into a cylindrical aluminum can, placed inside an orange cryostat. Data were collected at various temperatures in the 1.5-150 K range. Inelastic-neutron-scattering (INS) measurements were performed on the same powder sample using the HYSPEC spectrometer operated with the incident energies $E_i$ = 25 and 3.8 meV and a Fermi chopper frequency of 360 Hz (Supporting Information). The FullProf Software Suite was used to perform the structural and magnetic structures refinements.[58] Symmetry allowed magnetic structures were explored using representational analysis with the program *SARAh*[59] and magnetic symmetry analysis using magnetic space groups was performed using the Bilbao Crystallographic Server.[60-61] The INS data reduction and visualization were done with the MANTID[62] software Server Spin-wave calculations were performed using linear spin-wave theory as implemented in SpinW.[63]

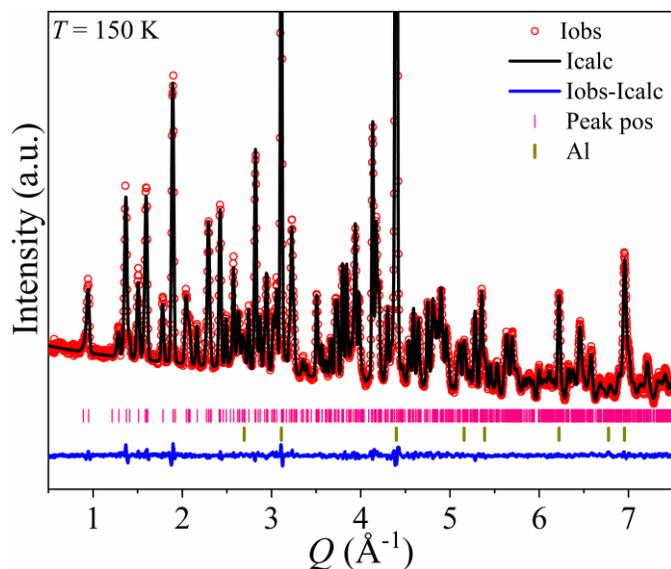

**Figure 1.** The Rietveld profile fit of the neutron powder diffraction data collected on HB2A using $\lambda = 1.54$ Å at 150 K. The refinement was performed using a structural model with orthorhombic symmetry in the *Pnma* (no.62) space group. No evidence of any impurity phases was detected within the instrument sensitivity. The refined parameters are $R_{wp}(\%) = 2.26$, $a = 13.2299(2)$ Å, $b = 6.02287(9)$ Å, $c = 8.32157(3)$ Å.

**Table 1.** Crystallographic data of NaCo$_2$(SeO$_3$)$_2$(OH) determined by single crystal X-ray diffraction.

| empirical formula | NaCo$_2$(SeO$_3$)$_2$(OH) |
|---|---|
| formula weight (g/mol) | 411.78 |
| crystal system | orthorhombic |
| space group, $Z$ | *Pnma* (no.62), *4* |
| Crystal dimensions, mm | 0.08 x 0.02 x 0.02 |
| *T*, K | 298 |
| *a*, Å | 13.2373(4) |
| *b*, Å | 6.0322(2) |
| *c*, Å | 8.3197(2) |
| volume, Å$^3$ | 664.33(3) |
| *D*(calc), g/cm$^3$ | 4.117 |
| $\mu$ (Mo K$\alpha$), mm$^{-1}$ | 16.007 |
| *F*(000) | 760 |
| *T*max, *T*min | 1.0000, 0.6887 |
| $2\theta$ range | 2.89-30.31 |
| reflections collected | 7327 |
| data/restraints/parameters | 659/1/74 |
| final *R* [*I*> 2σ(*I*)] $R_1$, $R_{w2}$ | 0.0138, 0.0349 |
| final *R* (all data) $R_1$, $R_{w2}$ | 0.0160, 0.0369 |
| GoF | 1.086 |
| largest diff. peak/hole, e/ Å$^3$ | 0.0380/-0.498 |

**Table 2.** Fractional atomic coordinates and isotropic displacement parameters (Å$^2$) of NaCo$_2$(SeO$_3$)$_2$(OH).

| Atom | *Wyckoff* | *x* | *y* | *z* | *Ueq* |
|---|---|---|---|---|---|
| Na(1) | 4*c* | 0.2515(1) | 0.2500 | -0.1288(2) | 0.0187(4) |
| Co(1) | 4*c* | 0.1129(4) | 0.2500 | 0.2124(7) | 0.0102(2) |
| Co(2) | 4*b* | 0.5000 | 0 | 0 | 0.0096(2) |
| Se(1) | 4*c* | 0.3292(3) | 0.2500 | 0.2251(5) | 0.0134(2) |
| Se(2) | 4*c* | -0.0249(3) | 0.2500 | -0.1195(5) | 0.0095(4) |
| O(1) | 4*c* | 0.3981(2) | 0.2500 | -0.0372(4) | 0.0109(4) |
| O(2) | 4*c* | 0.3912(2) | 0.2500 | 0.0431(3) | 0.0112(4) |
| O(3) | 4*c* | 0.0899(2) | 0.2500 | -0.0323(4) | 0.0172(4) |
| O(4) | 8*d* | -0.0146(1) | 0.0302(4) | -0.2470(2) | 0.0133(4) |
| O(5) | 8*d* | 0.2455(1) | 0.0413(4) | 0.1955(3) | 0.0194(5) |
| H(1) | 8*d* | 0.343(3) | 0.2500 | 0.029(6) | 0.048(4) |

**Table 3.** Selected bond distances (Å) and angles (°) of NaCo$_2$(SeO$_3$)$_2$(OH).

| Co(1)O$_6$ |  | Co(2)O$_6$ |  |
|---|---|---|---|
| Co(1)–O(1) | 2.008(3) | Co(2)–O(1) x 2 | 2.047(2) |
| Co(1)–O(3) | 2.058(3) | Co(2)–O(2) x 2 | 2.116(2) |
| Co(1)–O(4) x 2 | 2.153(2) | Co(2)–O(4) x 2 | 2.121(2) |
| Co(1)–O(5) x 2 | 2.165(2) |  |  |
| **Se(1)O$_3$** |  | **Se(2)O$_3$** |  |
| Se(1)–O(2) | 1.722(3) | Se(2)–O(3) | 1.685(3) |
| Se(1)–O(5) x 2 | 1.695(2) | Se(2)–O(4) x 2 | 1.703(2) |
|  |  |  |  |
| Co(1)–O(1)–Co(2) | 101.38(1) | Co(1)⋯Co(2) | 3.199(5) |
| Co(1)–O(4)–Co(2) | 96.92(1) | Co(2)⋯Co(2) | 3.016(3) |
| Co(2)–O(1)–Co(2) | 94.93(1) |  |  |
| Co(2)–O(2)–Co(2) | 90.90(1) |  |  |

## 3. Results

### 3.1. Crystal structure of NaCo$_2$(SeO$_3$)$_2$(OH)

The obtained columnar NaCo$_2$(SeO$_3$)$_2$(OH) crystals with an average length of ~0.2 mm were used for SXRD measurements (Figure SI1). The SXRD analysis confirms that NaCo$_2$(SeO$_3$)$_2$(OH) crystallizes in orthorhombic crystal system with space group of *Pnma* (No. 62) with unit cell parameters $a$ = 13.2373(4) Å, $b$ = 6.0322(2) Å, $c$ = 8.3197(2), $V$ = 664.33(3) Å$^3$. The detailed crystallographic data are reported in Table 1-2. The structure contains one symmetry unique Na$^+$, two Co$^{2+}$, two Se$^{4+}$, five O$^{2-}$ and one H$^+$ ions. The two Co$^{2+}$ are in 4$c$ (0.1129(3), 0.2500, 0.2124(7)) and 4$b$ (0.5, 0, 0) sites respectively, and form CoO$_6$-octahedra (*oct*). Two Se-sites (4$c$-sites) bind with three oxygen atoms by forming [SeO$_3$] units. A projection of the structure of NaCo$_2$(SeO$_3$)$_2$(OH) that displays the stacking of Co–O–Co sawtooth chain in the crystallographic *ac*-plane is shown in Figure 2a. The sawtooth chains are interconnected via [SeO$_3$] groups along the *a*- and *c*-axes. Moreover, along the *ab*-plane the sawtooth chains pack in a zigzag fashion while changing the pattern between nearest neighbor chains (Figure SI3). Compared to other sawtooth oxyanion compounds, NaCo$_2$(SeO$_3$)$_2$(OH) possess an overall 3D structure made from CoO$_6$-*oct* and [SeO$_3$]-units. As seen in Figure 2b, the Co(1)O$_6$-*oct* share edges with two Co(2)O$_6$-*oct* to form the triangular units in the sawtooth chain. These triangular units run up and down along the *b*-axis forming the sawtooth chain. Within each Co$_3$–triangle, one Co(1)O$_6$-*oct* share edges with two Co(2)O$_6$-*oct* via O(1), O(3) and O(5) and can be best described as [Co$_3$O$_{13}$] triangular units with O(1) serving as the $\mu_3$-oxo vertex which also binds to H by forming the only $^-$OH groups in the structure, Figure 2b.

A summary of the Co–O, Se–O bond distances and Co–O–Co bond angles are given in Table 3. The six coordinate CoO$_6$ units in NaCo$_2$(SeO$_3$)$_2$(OH) possess an average Co–O distance of 2.105(3) Å, which is comparable to the expected sum of the Shannon crystal radii, 2.145 Å, for a 6-coordinate high spin Co$^{2+}$ and O$^{2-}$.[64] The $\mu_3$-oxo bonds Co(1)–O(1) and Co(2)–O(1) are 2.008(3) and 2.047(2) Å, respectively and the resulting Co−O−Co bond angles for the edge-sharing connections of the CoO$_6$ groups range from 90.90(1)° to 101.38(1)°. The Co−O interatomic distances range from 2.008(3) Å to 2.165(2) Å of Co(1)O$_6$-*oct* and 2.047(3) Å to 2.121(2) Å of Co(2)O$_6$-*oct*, respectively indicating a high degree of distortion. The Co(1)–Co(2) ($J_{bv}$ coupling) and the Co(2)–Co(2) ($J_{bb}$ coupling) distances are 3.199(5) and 3.016(3) Å, respectively, as highlighted in Figure 2c. The bond angles of Co(2)–O(1)–Co(2) and Co(2)–O(2)–Co(2) of the base-base super exchange interaction pathways are 94.93(1)° and 90.90(1)°, respectively, while the base-vertex angles of Co(1)–O(1)–Co(2) and Co(1)–O(4)–Co(2) are 101.38(1)° and 96.92(5)°, respectively. The sawtooth chains are only ~5.5 Å apart, which is close enough to potentially allowing super-super exchange interactions between the chains. Therefore, in addition to super-exchange pathways via Co–O–Co connectivity in sawtooth chains, super-super exchange pathways via [SeO$_3$] groups are possible, potentially leading to complex magnetic properties due to competing interchain interactions.

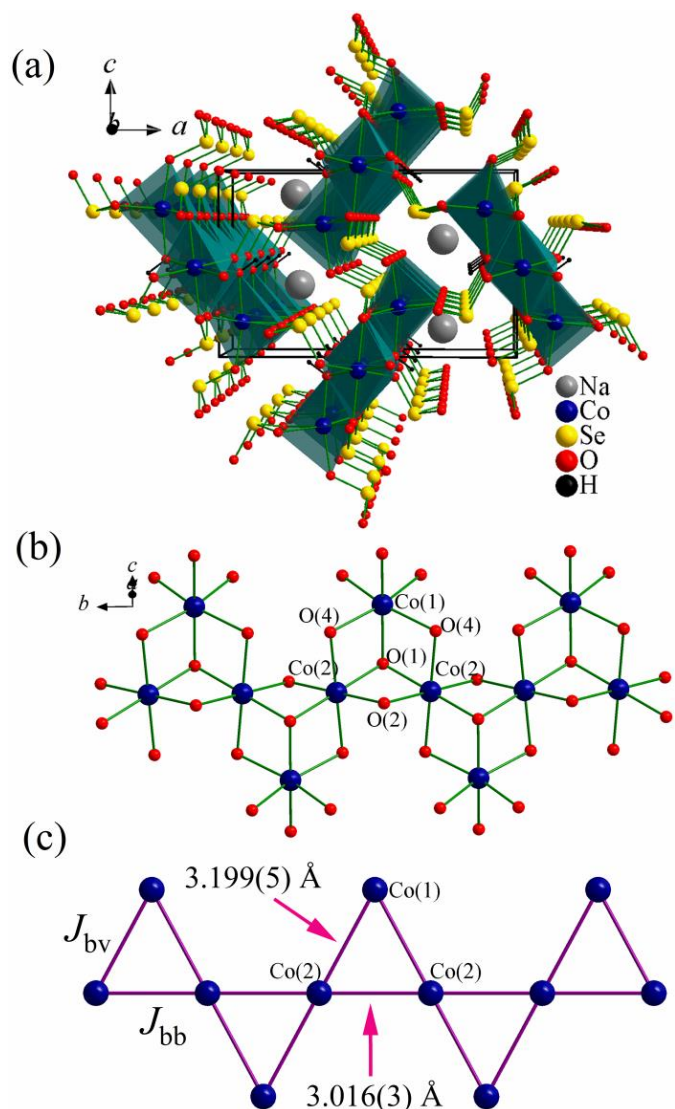

**Figure 2.** (a) Partial polyhedral view of NaCo$_2$(SeO$_3$)$_2$(OH) projected along the *b*-axis, showing packing of Co–O–Co sawtooth chains on the *ac*-plane. The Na$^+$ ions reside inside the channel structure which propagate along the *b*-axis. (b) Partial structure of Co–O–Co sawtooth chains made from edged sharing CoO$_6$ octahedra along the *b*-axis. (c) The unequal base-base ($J_{bb}$) and base-vertex ($J_{bv}$) exchange interactions are shown using the solid purple line. The Co(1)–Co(2) and the Co(2)–Co(2) distances are 3.199(5) and 3.016(3) Å, respectively.

### 3.2 Magnetic Properties of NaCo$_2$(SeO$_3$)$_2$(OH)

The temperature dependent magnetization, $\chi = M/H$ of NaCo$_2$(SeO$_3$)$_2$(OH) was measured on both zero-field cooling (ZFC) and field cooling (FC) modes between 2-350 K using ground single crystals. The magnetic susceptibility measured with an applied magnetic field of 10 kOe is shown in Figure 3a which exhibits a sharp rise below 20 K and reaches to a maximum at ~11 K ($T_1$) suggesting a ferromagnetic-type transition. The inverse magnetic susceptibility in the paramagnetic region for applied magnetic field $H = 10$ kOe can be fit using the Curie-Weiss model $M/H = C (T-\theta)$. As shown in Figure 3a, the fit above 150 K resulted an effective moment of 4.84 $\mu_B$/Co and a Weiss temperature of -9 K. The effective magnetic moment is higher than the spin only value of Co$^{2+}$ with $S = 3/2$, (i.e. $\mu_{eff} = 3.8\ \mu_B$). This could be due to the significant orbital contribution since orbital moments are unquenched for Co$^{2+}$ in octahedral environment ($t_{2g}^5 e_g^2$, $S = 3/2$, $L = 3$). The negative $\theta_{cw}$ confirms the overall antiferromagnetic nature of NaCo$_2$(SeO$_3$)$_2$(OH). It is noteworthy that the frustration index $f = |\theta_w/T_N|$ is ~1 even though Co(1) and Co(2) form a geometrically frustrated triangular lattice. This frustration index is very small compared to our previously reported sawtooth chain systems, Rb$_2$Fe$_2$O(AsO$_4$)$_2$ ($f = $ ~20)[32] and Rb$_2$Mn$_3$(MoO$_4$)$_3$(OH)$_2$ ($f = $ ~24).[12]

Figure 3b shows the temperature dependence of the magnetic susceptibility ($\chi = M/H$) for $T < 25$ K collected under various magnetic fields ($H = 100$ Oe, 1 kOe, 10 kOe, 30 kOe and 50 kOe) for both ZFC and FC measurement protocols focusing on the low temperature region. For a 100 Oe applied magnetic field, the FC and ZFC curves bifurcate at 11 K ($T_1$), each exhibiting a broad hump before starting to go down at 6 K. Below 6 K, the magnetic susceptibility continuously drops with lowering temperature and at 2 K it reaches very close to zero. This downturn in the magnetic susceptibility at 6 K can be identified as a second magnetic transition, $T_2 = 6$ K. This is further confirmed by heat capacity and neutron diffraction experiments as will be discussed in later sections. Furthermore, increasing the magnetic field above 100 Oe suppresses the magnitude of this plateau-like feature and broadens it pushing the second transition to lower temperatures as well as decreasing the separation between FC and ZFC measurements. Finally, at an applied field of 10 kOe the bifurcation between FC and ZFC measurements disappears. As displayed in Figure 3b inset, at applied magnetic fields of 30 and 50 kOe, the magnetic susceptibility starts to increase below 6 K potentially suggesting another field induced transition near $T_2$. This was clearly observed in in our heat capacity measurements. It is important to mention that the occurrence of two successive magnetic transitions in transition metal-based oxy-anion sawtooth compounds is very rare. Similar consecutive transitions were observed in one of our half sawtooth compounds, $Rb_2Mn_3(MoO_4)_3(OH)_2$ where it shows a transition from a paramagnetic to an incommensurate phase below 4.5 K followed by another commensurate antiferromagnetic transition below 3.5 K.[12]

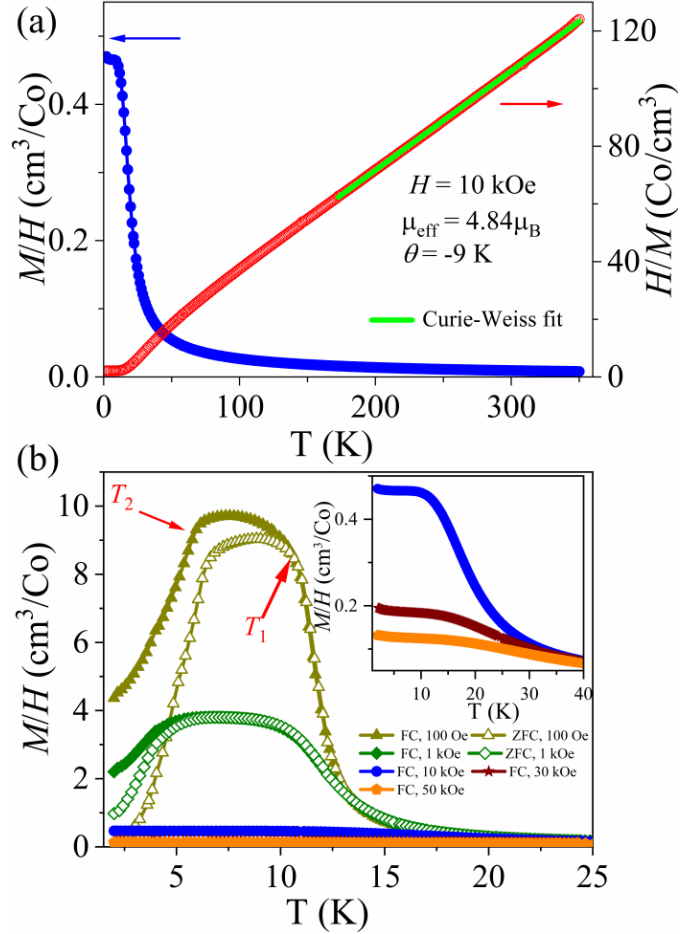

**Figure 3.** Magnetic susceptibility, $\chi$, and inverse magnetic susceptibility, $1/\chi$, as a function of temperature, $T$, measured using ground single crystal sample of $NaCo_2(SeO_3)_2(OH)$ measured at 10 kOe. (b) Magnetization curves of $NaCo_2(SeO_3)_2(OH)$ measured in different magnetic fields ($H$ = 100 Oe – 50 kOe) below 25 K.

Figure 4a shows the isothermal magnetization measured at various temperatures, $T$ = 2, 4, 5, 10, 15, 20 and 100 K up to 60 kOe. At 2 K, the isothermal magnetization curve exhibits hysteresis which demonstrates the presence of a ferromagnetic component to the magnetic structure, Figure 4b. In addition to a hysteresis in the ascending and descending magnetization curves, an anomaly can be seen at $H_c$ = 1.3 kOe which is likely related to a spin flip transition. This field induced transition at 1.3 kOe agrees with our magnetic susceptibility data as shown in the inset of Figure 3b where the magnetic susceptibility peak starting to disappear at 1 kOe. After the field induced transition, the magnetization follows a concave curvature with increasing field and reaches a maximum of 1.2 μ$_B$/Co at 60 kOe which is smaller than the fully saturated moment of high-spin Co$^{2+}$. The field induced transition and the concave nature of the magnetization curve possibly indicate a canted AFM phase or a partially polarized Co-spin state in NaCo$_2$(SeO$_3$)$_2$(OH). The field induced transition at $H_c$ = 1.3 kOe disappears with increasing temperature however, the magnetization curves exhibit a curvature behavior even beyond $T_1$ as displayed in Figure 4a.

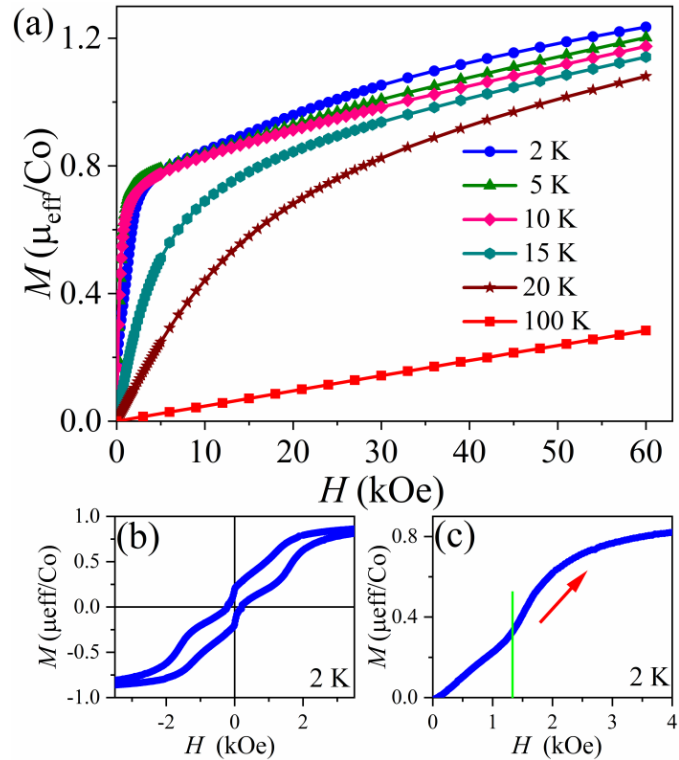

**Figure 4.** Isothermal magnetization data obtained at 2–100 K up to 60 kOe. (b) A blow-up region of the hysteresis at 2 K with ascending and descending magnetization curves. (c) The magnetization data measure at 2 K with field increasing which indicates the field induced transition at 1.3 kOe.

### *3.2 Heat Capacity of NaCo$_2$(SeO$_3$)$_2$(OH)*

The temperature dependent heat capacity of NaCo$_2$(SeO$_3$)$_2$(OH) was measured in applied fields of up to 110 kOe using a 2.6 mg pressed pellet, as shown in Figure 5. At zero applied magnetic field a sharp λ-anomaly peak at $T_1$ = 11 K was observed which agrees well with our magnetic susceptibility data (Figure 3). Interestingly, another broader transition was observed below 6 K which reaches to a maximum at about 3.8 K (Figure 5 inset). As displayed in Figure 5, applied magnetic fields of magnitudes less than $H_c$ has very little effect on both the magnitude and the peak position in heat capacity at $T_1$. With further increasing magnetic field, the λ-anomaly becomes much broader over a wide range of temperatures. Apart from this field induced transition at $T_1$, a very interesting and complex field dependent behavior was observed at $T_2$ up to 110 kOe. Starting from 30 kOe applied magnetic field, a small peak appeared at 2.5 K which starts to grow with applied magnetic field up to 110 kOe. First, this new peak moves to higher temperatures at an applied magnetic field of 50 kOe, then it starts to move to lower temperatures at about 90 kOe indicating a complex magnetic phase diagram with non-monotonic field dependence. This field induced transition was also observed in our magnetization data, manifesting as an increase in the magnetic susceptibility at $T_2$, as shown in Figure 3b inset. Neutron powder or single crystal diffraction in applied magnetic fields will be necessary to understand the field induced magnetic ground state at $T_2$.

To calculate the magnetic entropy, the heat capacity data of $NaCo_2(SeO_3)_2(OH)$ above 30 K were fitted by an Einstein+Debye model as the lattice contribution ($C_l$). The magnetic contribution, $C_m$, was calculated as the difference of $C_p$-$C_l$, and the magnetic entropy, $S_M$, was estimated by integrating $C_{mag}/TdT$. As displayed in the inset of Figure 5, the calculated magnetic entropy is $S_M$ = 5.92 J mol$^-$ K$^-$, representing 51% from the entropy $\Delta S = R\ln(2S+1) = 11.5$ J mol$^-$ K$^-$ for $Co^{2+}$ in $S = 3/2$ spin state, but just slightly larger than what is expected for an effective spin $S_{eff} = ½$, $R\ln2$ = 5.76 mol$^-$ K$^-$. However, a more careful lattice contribution subtraction using a nonmagnetic analogue of $NaCo_2(SeO_3)_2(OH)$ such as $NaZn_2(SeO_3)_2(OH)$ is necessary to increase the confidence of the entropy value.

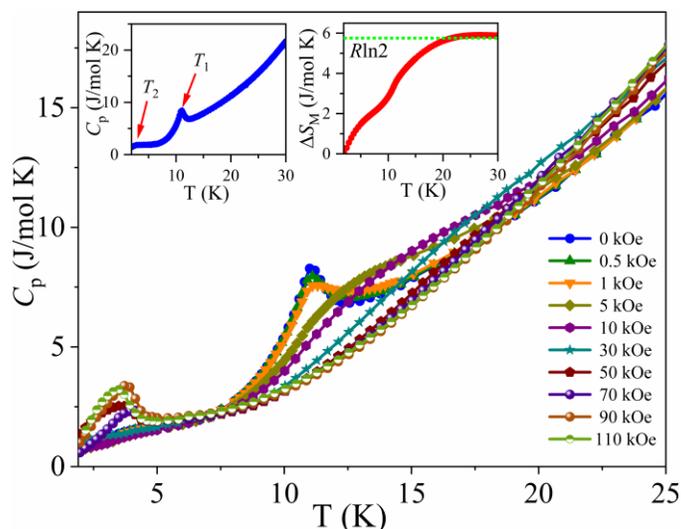

**Figure 5.** The specific heat curves of NaCo$_2$(SeO$_3$)$_2$(OH) obtained for applied magnetic fields in 0 – 110 kOe range. A significant influence of the external magnetic field on both $T_1$ and $T_2$ transitions is observed. The insets show a zoomed view of the two magnetic transitions measured at $H = 0$ T, and the evaluated magnetic entropy of NaCo$_2$(SeO$_3$)$_2$(OH), respectively.

### 3.5 Magnetic Structure of NaCo$_2$(SeO$_3$)$_2$(OH)

Neutron powder diffraction was performed using the HB-2A powder diffractometer at the High Flux Isotope Reactor, Oak Ridge National Laboratory.[57] The crystal structure of NaCo$_2$(SeO$_3$)$_2$(OH) was confirmed by performing Rietveld refinements using neutron powder diffraction data collected above the signals observed in the bulk probes at 150 K (Figure 1). Here, data was collected using the 1.54 Å wavelength to cover a broader range of scattering angle and collect more nuclear peaks for use in the nuclear refinement with the Ge(115) monochromator. (the refined parameters are available in Table SI1). Figure 6a displays NPD data collected at 30, 6 and 1.5 K. As shown additional peaks are seen in both the 6 and 1.5 K data consistent with both the $T_1$ and $T_2$ transitions identified in the bulk measurements indicating two distinct successive magnetic transitions. We note that for the 6 K data, due to the temperature being too close to the second transition temperature ($T_2$) the NPD collected at 6 K also includes a small contribution from the lower temperature magnetic phase. However, as shown in order parameter scans in Figure 6b the two sets of peaks clearly display different temperature dependencies and so are attributable to separate phase transitions. The first magnetic transition ($T_1$) produces magnetic reflections on positions indexed by the nuclear unit cell, thus indicating an ordering vector ($k$) of $k = (0, 0, 0)$. However, additional magnetic peaks are observed in the 1.5 K data that can be indexed by the wave vector $k = (1/2, 0, 0)$. Figure 6b presents the temperature dependence of the (101) Bragg peak intensity (peak at $Q\sim0.89$ Å$^{-1}$). On cooling, the (101) peak intensity first increases at 11 K starting from a background equivalent count rate until it reaches a maximum at 8.5 K, and then quickly decreases at ~6 K down to a constant value that corresponds to roughly one third of its maximum value. On the other hand, the (3/2, 0, 0) magnetic peak (at $Q\sim0.71$ Å$^{-1}$) appears at about 6 K and increases until it saturates at about 3 K. These order parameter measurements agree with the magnetic and heat capacity data.

To solve the magnetic structure of the first ordered state ($T_1 < T < T_2$), we analyzed the NPD measured at 6 K. The Bilbao Crystallographic server was utilized to find the maximal magnetic space groups allowed from the parent *Pnma* crystal structure and the $k_1 = (0, 0, 0)$ propagation vector. Of the allowed magnetic space groups (MSG) our analysis found *Pn'ma'* (#62.448) to provide the best fit to the observed diffraction pattern. According to our model, at $T_1$ only the Co(2) site orders while the Co(1) stays paramagnetic, Figure 7. The refined magnetic structure reveals that the Co(2) spin orders in a canted ferromagnetic structure with the net magnetization aligned parallel to the *b*-axis, Figure 7a top panel. The ordered component along the *b*-axis is aligned ferromagnetically both within the sawtooth chain and between the neighboring chains. In contrast, the smaller components along the *a* and *c* directions are alternating their directions both within and between the chains leading to no net moment in these directions. The refined magnetic moment components are $m_a = 0.72(4)$ μ$_B$, $m_b = 2.56(1)$ μ$_B$ and $m_c = 0.45(3)$ μ$_B$ for Co(2) and yield a total static moment of 2.7μ$_B$. This is less than the fully ordered spin only value of Co$^{2+}$ with $S = 3/2$, $\mu_{eff} = 3.8$ μ$_B$.

With a model for the higher temperature structure we turn now our attention to the additional magnetic reflections corresponding to $k_2 = (1/2, 0, 0)$ that were seen to arise below 6 K, noting that the magnetic peaks corresponding to $k_1 = (0, 0, 0)$ remain present but weaken in intensity. Of the allowed magnetic symmetries compatible with both wavevectors ($k_1$, $k_2$), the model based on *Pn'a2$_1$'* (#33.146) space group in a unit cell of (2a,b,c) is best able to model the 1.5 K data. In this model Co(1) and Co(2) moments order with total magnitude of 1.32 μ$_B$ ($m_a$ = 0.60(1) μ$_B$, $m_b$ = -0.2(1) μ$_B$, $m_c$ = 1.15(4) μ$_B$ ) and 2.97 μ$_B$ ($m_a$ = 2.09(3) μ$_B$, $m_b$ = 1.46(1) μ$_B$, $m_c$ = 1.53(1) μ$_B$), respectively. As such, the magnitude of the Co(1) moment that only orders below $T_2$ is significantly smaller than the Co(2) moment. We also note that the $m_b$ components are arranged in ferromagnetic manner, following the $k_1$ = (0,0,0) wavevector, while the $m_a$ and $m_c$ components are arranges antiferromagnetically according to the $k_2$ = (1/2, 0, 0) propagation vector for both sites. The negative sign of the $m_b$ for Co(1) denotes an opposite alignment of this component to the $m_b$ of the Co(2) leading to ferrimagnetism along the *b*-axis. A graphical representation of the proposed magnetic structure is given in Figure 7 bottom panels. The Co(2) moments feature a colinear ferromagnetic arrangement inside the chain and are canted away from the plane of the triangular chain. The Co(2) moments from neighboring chains have a *b*-axis component aligned ferromagnetically while the *a*- and *c*- components are fully compensated. In contrast, the Co(1) magnetic moments lies in the plane of the Co–O–Co triangles (Figure 7b bottom panel), with the predominant *a*- and *c*- components compensated among the neighboring chains. This produces a dominant AFM nature to the NaCo$_2$(SeO$_3$)$_2$(OH) below $T_2$. This behavior appears consistent with our magnetization data and the heat capacity, however, the underlying force of these two magnetic transitions is still unclear. Considering the stability of the two separate magnetic structures, we could suggest that at lower temperatures next nearest neighbor (NNN) interactions become stronger than the nearest neighbor (NN) interaction driving the overall magnetic structure to an AFM structure. As previously stated, within the triangular [$Co_3O_{13}$] motif the NN Co–O–Co angles range from 90.90(1)° to 101.38(1)° with NN distances of 3.199(5), $J_{bv}$ and 3.016(3) Å, $J_{bb}$. Since, the bond angles are close to 90°, a dominant FM interaction can be expected within the Co–O–Co sawtooth chain according to the Goodenough Kanamori rule.[65-66] In the overall structure, the individual Co–O–Co chains are interconnected via [$SeO_3$] groups to form a 3D structure. Therefore, Co–O–Se–O–Co interactions (NNN) are likely relevant for the magnetic behavior at lower temperature.

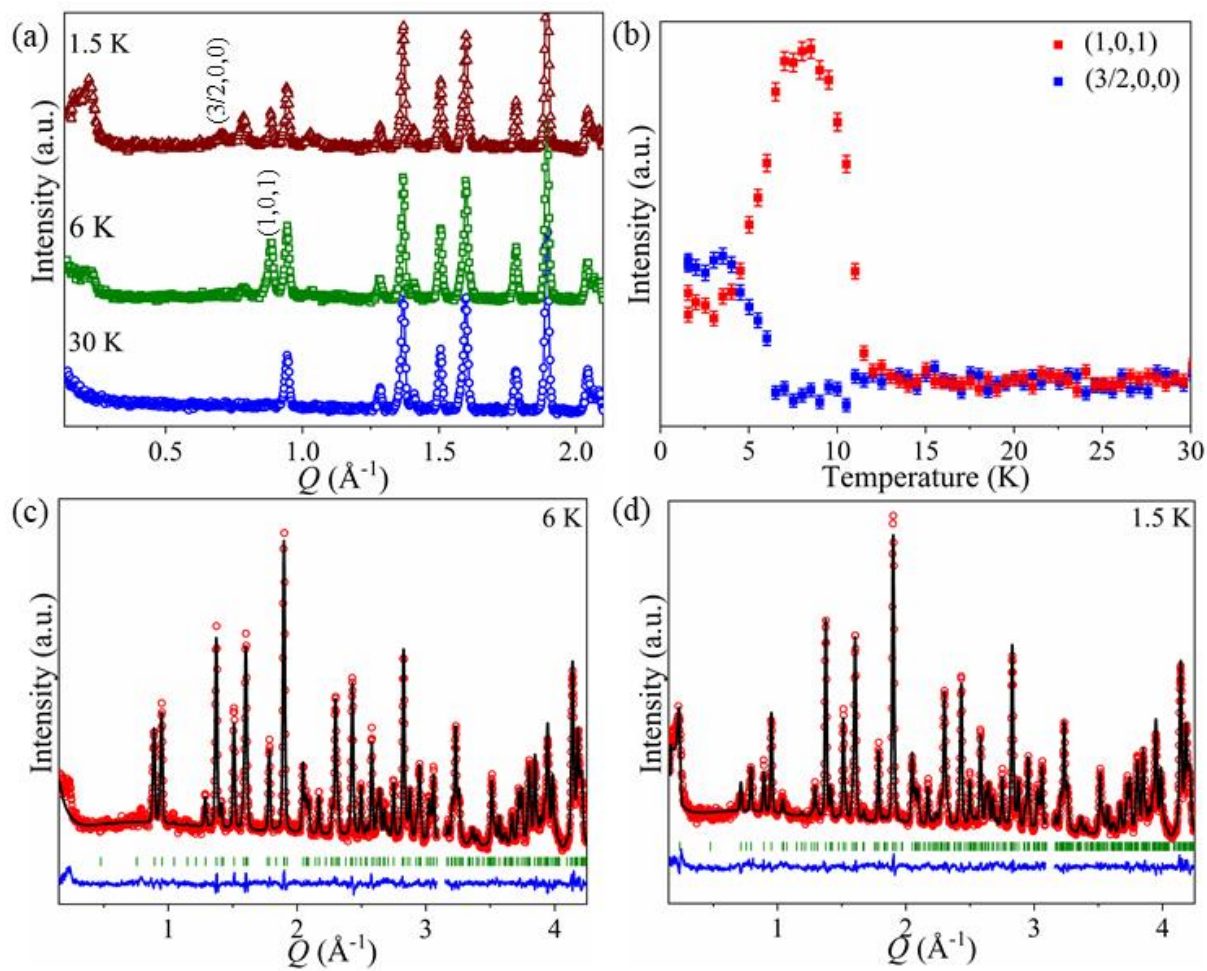

**Figure 6.** (a) Neutron powder diffraction patterns of NaCo$_2$(SeO$_3$)$_2$(OH) at 30, 6 and 1.5 K showing the appearance of new peaks below $T_1$ and $T_2$ transition temperatures. (b) The intensities of (1,0,1) and (3/2,0,0) peaks as a function of temperature. NPD data refinements at 6 K (c) and 1.5 K (d). In both panels the data, model, difference, and phase peak indexes are indicated by red circle, black line, blue line, and green tick marks, respectively. The gaps in the patterns are due to regions excluded to remove strong Al peaks from the sample can.

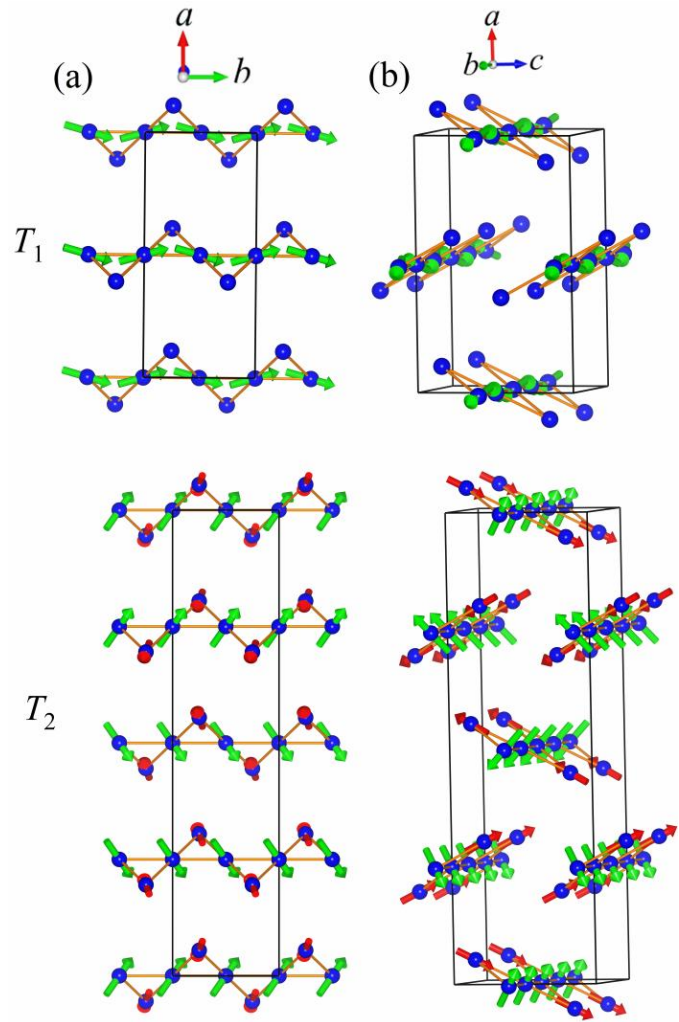

**Figure 7.** The proposed magnetic models for the two magnetic transitions, 11 K ($T_1$) and 6 K ($T_2$) are visualized in two different crystallographic orientations. The magnetic structure at 11 K is described within the same unit cell as the nuclear structure, whereas the structure at 6 K is doubled along the *a*-axis (i.e. 2*a, b, c*).

### 3.6 Crystal field Excitations of NaCo$_2$(SeO$_3$)$_2$(OH)

Neutron inelastic scattering data on polycrystalline NaCo$_2$(SeO$_3$)$_2$(OH) measured at 1.5 K and 50 K with an incident energy, $E_i$ = 25 meV is displayed in Figure 8. The spectrum consists of two flat excitation bands centered at about at 7 and 17 meV, which broaden progressively as temperature increases. The |**Q**|-integrated cuts over the range 1 - 2.5 Å$^{-1}$, shown in Figure 8c, suggest that each band is structured into at least two distinct excitations within the experimental resolution. The instrumental energy resolution (the full width at half maximum) is represented by vertical blue lines in Figure 4c. The decay of intensity with the momentum transfer as well as the temperature dependence demonstrate the magnetic nature of the excitations. The presence of pair excitations in each band is not surprising considering the existence of two distinct Co$^{2+}$ magnetic sites in our material.

The $d^7$ electrons of Co$^{2+}$ in an octahedral crystal field can possess a multiplet state with spin $S$ = 3/2 and effective orbital moment $L_{eff}$ = 1. The spin-orbit coupling (H$_{SO}$ = $\lambda \vec{L} \cdot \vec{S}$) can split this state into three states: a $J_{eff}$ = 1/2 ground state, and $J_{eff}$ = 3/2 and $J_{eff}$ = 5/2 excited states, that are separated in energy by 3/2 $\lambda$ and 5/2 $\lambda$, respectively. The observation of crystal field excitations between the $J_{eff}$ = 1/2 ground state and the excited states is common in cobalt compounds and provides support for the presence of the spin-orbital entangled state.[66-71] The deviation from ideal octahedral environment produces a further split of the $J_{eff}$ = 3/2 and 5/2 excited states. Moreover, the long-range magnetic order below $T_N$ creates an internal molecular field, which induces a Zeeman splitting of the $J_{eff}$ = 1/2 manifold.

The inelastic spectrum of NaCo$_2$(SeO$_3$)$_2$(OH) is reminiscent to the spectra observed for α-CoV$_2$O$_6$ and α, γ-CoV$_3$O$_8$ that show two bands of excitations at ~ 5 meV and ~ 25 meV, which are due to transitions within the $J_{eff}$ = 1/2 manifold and between the $J_{eff}$ = 1/2 and $J_{eff}$ = 3/2 manifolds, respectively.[66-68] Similarly, we associate the flat band at 17 meV of NaCo$_2$(SeO$_3$)$_2$(OH) with the spin-orbit excitations between the $J_{eff}$ = 1/2 and $J_{eff}$ = 3/2 manifolds in the two distinct Co$^{2+}$ sites. Furthermore, the fact that the intensity of the 7 meV excitation is significantly reduced at 50 K, above the magnetic order transition, suggests that this excitation is associated with the splitting of the lowest-energy ground-state $J_{eff}$ = 1/2 doublet due to the molecular field induced by magnetic order. Hence, the observed spin-orbit excitations support the view that Co$^{2+}$ ions of NaCo$_2$(SeO$_3$)$_2$(OH) have a spin-orbital entangled $J_{eff}$ = 1/2 state.

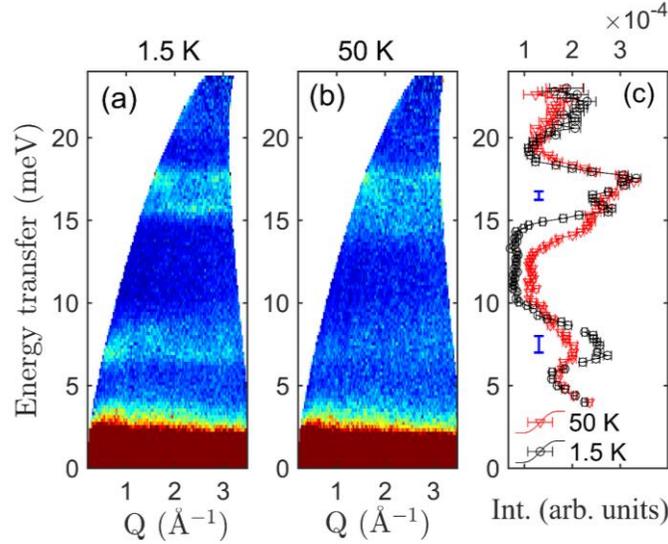

**Figure 8.** Inelastic neutron scattering data obtained with incident energy $E_i$ = 25 meV at two temperatures (a) 1.5 K and (c) 50 K. (c) |**Q**|-integrated cuts (|**Q**| = [1 - 2.5] Å$^{-1}$) of inelastic scattering showing the presence of two bands of crystal field excitations. Vertical blue lines in (c) indicate the instrumental energy resolution (full width at half-maximum).

### 3.7 Spin-wave excitations of NaCo$_2$(SeO$_3$)$_2$(OH)

The low-energy excitation spectrum of NaCo$_2$(SeO$_3$)$_2$(OH) was characterized using the HYSPEC spectrometer with the incident energy $E_i$ = 3.8 meV. At low temperature ($T$ = 1.5 K), the powdered averaged spectrum of the magnetically ordered state, shown in Figure 9a, features two well-defined excitation modes: one flat mode located at about 1.1 meV and a second dispersive mode that emerges at the momentum transfer $Q \approx 0.8$ Å$^{-1}$, corresponding to the (3/2, 0, 1) magnetic peak, and extends up to about 0.6 meV. At 8 K, in the partially ordered state, $T_1 < T < T_2$, both excitation modes vanish, and only diffuse scattering is observed, Figure 9b.

An attempt of describing the spin-wave spectrum was made using a Heisenberg Hamiltonian model which includes three intrachain and two interchain exchange parameters, as depicted in Figure 9d. For the intrachain couplings we considered the exchange interactions between the nearest-neighbor Co(2) atoms forming the triangles base ($J_{bb}$), the interaction between the base and vertex positions Co(1)–Co(2) ($J_{bv}$), as well as the next-nearest neighbor base-vertex interaction ($J_{bv2}$), Figure 9d. The long-range magnetic order is stabilized by additional couplings between adjacent sawtooth chains ($J_{n1}$ and $J_{n2}$), that involve super-exchange pathways via [SeO$_3$] groups (Co(1)–O–Se–O–Co(1) or Co(1)–O–Se–O–Co(2)). The spin-wave spectrum was calculated using linear spin wave theory as implemented in the SpinW program. The analyses indicate that the main characteristics of excitation spectrum can be reproduced using the following parameters: $S \cdot J_{bb}$ =

0.15 meV, $S \cdot J_{bv}$ = 0.4 meV, $S \cdot J_{bv2}$ = 0.1 meV, $S \cdot J_{n1}$ = -0.05 meV, $S \cdot J_{n2}$ = -0.03 meV. The positive $J$ values correspond to antiferromagnetic coupling while the negative represent a ferromagnetic coupling. We found that the ferromagnet interchain couplings ($J_{n1}$ and $J_{n2}$) play a key role in stabilizing the AFM order with the wavevector $\boldsymbol{k}$ = (1/2, 0, 0). The large ratio $J_{bv}/J_{bb}$ could be due to the difference in Co–O–Co bond angles associated with the two exchange pathways. The Co(1)–O–Co(2) angles ($J_{bv}$) are ~ 91° and 95°, while the Co(2)–O–Co(2) ($J_{bb}$) angles are 97° and 100.5°. This simple Heisenberg Hamiltonian model is successful in reproducing the energy positions of two spin-wave modes but is deficient in describing the correct $\boldsymbol{Q}$ dependence of the intensity for the low-energy mode (see Figure 9c). Moreover, the model fails in explaining the ferromagnetic components along $b$-direction, as well as the orthogonal arrangement between Co(1) and Co(2) moments. Considering the significant information loss in the powder averaged inelastic data, the number of parameters that can be tested simultaneously is drastically limited. This precludes the use of more complex models that include further neighbor interactions, biquadratic coupling or anisotropic exchange interactions that are expected to arise from the strong spin-orbit coupling in this compound. However, additional spin wave excitation studies using single crystals are needed to look for firmer quantitative evidence of intrachain and interchain exchange parameters of $NaCo_2(SeO_3)_2(OH)$.

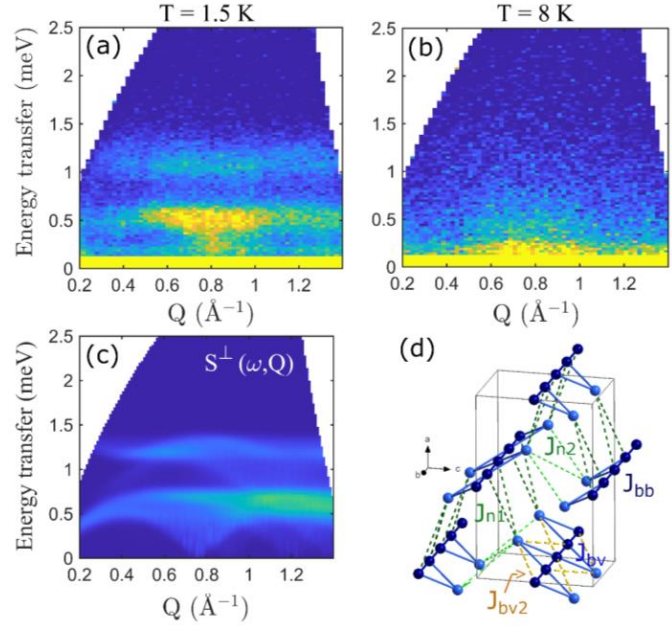

**Figure 9.** (a) Spin-wave excitation spectrum of measured at 1.5 K using $E_i$ = 3.8 meV. (b) Inelastic spectrum measured in the partial ordered state at 8 K, ($T_1 < T < T_2$) (c) Calculated spin-wave spectrum using a simple Heisenberg Hamiltonian model that includes three intra-chain ($J_{bv}$, $J_{bb}$, $J_{bv2}$) and two interchain ($J_{n1}$, $J_{n2}$) exchange interactions. The magnetic exchange pathways are shown in (d).

## 4. Conclusions

Novel magnetic materials with $Co^{2+}$ are promising candidates to study emergent physics since unquenched orbital moments and large degeneracy of the orbitals could facilitate magnetic frustration and anisotropy. The stereo active asymmetric selenite [$SeO_3$] group and $CoO_6$-octahedra can stabilize promising magnetic materials with enormous range of chemical bonding modes. In this study we have examined a novel delta chain compound, $NaCo_2(SeO_3)_2(OH)$ through a combination of bulk magnetization measurements, neutron powder diffraction and inelastic neutron scattering. Single crystals of $NaCo_2(SeO_3)_2(OH)$ was synthesized using a hydrothermal method. $NaCo_2(SeO_3)_2(OH)$ consists of two crystallographically distinct $Co^{2+}$ sites (Co(1) and Co(2)). These two distinct sites form sawtooth triangular 1D chains that propagate along the *b*-axis. The chains are interconnected via non-magnetic [$SeO_3$] groups producing a 3D structure. Due to the presence of two different crystallographic sites, two different nearest neighbor interactions form within the triangle between the *base-base* and *base-vertex* Co sites. Additionally, multiple next nearest neighbor interactions are present via non-magnetic linker group, [SeO$_3$].

Our macroscopic magnetic measurements and neutron powder diffraction data confirm that there are two successive magnetic phase transition on cooling. The first transition is from a paramagnetic phase to a canted FM phase at $T_1$ = 11 K, described by a propagation vector of $k_1$ = (0, 0, 0). A second magnetic transition occurs below $T_2$ = 6 K going from FM to a canted AFM structure that is defined by two $k$ vectors: $k_1$ = (0, 0, 0) and $k_2$ = (1/2, 0, 0). Our proposed magnetic structure model suggest that at 11 K, only the Co(2) site is magnetically ordered in an alternating canted configuration about the *b*-axis, while the Co(1) site stays disordered. Below 6 K both Co(1) and Co(2) order and form a canted AFM structure. Within the delta chain, Co(1) and Co(2) form a complex spin topology with the Co(1) moment lying inside the sawtooth triangular plane and the Co(2) spin canted out of the plane. These magnetic moments are fully compensated in the *ac*-plane but remain uncompensated along the chain axis (*b*-direction). The ground state of Co$^{2+}$ was elucidated by measuring the crystal field excitations using inelastic neutron scattering, which supports the presence of $J_{\text{eff}}$ =1/2 state due to the unquenched spin-orbital coupling. Crystal field excitation clearly reveals two transitions at 7 and 17 meV. The 17 meV transition is attributed to the spin-orbit excitation between $J_{\text{eff}}$ = 1/2 and $J_{\text{eff}}$ = 3/2 states that characterize both Co$^{2+}$ sites. The temperature dependence of the scattering suggests that 7 meV excitation is associated with the splitting of the $J_{\text{eff}}$ = 1/2 doublet due to the molecular field induced by magnetic order. Low-energy powder inelastic measurements revealed the presence of spin-wave excitations associated to the lower temperature magnetically ordered state. The spin-wave spectrum contains a dispersive excitation that extends up to about 0.6 meV and a nearly flat excitation mode centered at 1.1 meV.

A Heisenberg Hamiltonian model that includes three intrachain exchange interactions and two super-exchange interchain couplings is insufficient to describe the sequence in magnetic ordering and the complex canted antiferromagnetic ground state. A more complete understanding of what leads to these competing magnetic interactions would require a careful inelastic neutron scattering experiment using aligned single crystals and we leave this to future neutron scattering work.

To conclude, our work highlights the complexity of frustrated magnetism emerging from weakly coupled sawtooth chain in the $Co^{2+}$-based compound $NaCo_2(SeO_3)_2(OH)$. Further single crystal studies are necessary to develop a full understanding of this system. We hope that our study will inspire the search for other $Co^{2+}$-based sawtooth oxyanions structures based on phosphate, arsenates and molybdates targeting new physics and potential applications. Collectively our results highlight the complexity of magnetism in sawtooth structures.

**ASSOCIATED CONTENT:**

**Supporting Information**

The Supporting Information include crystallographic refinement data from the neutron powder diffraction of $NaCo_2(SeO_3)_2(OH)$ at 150 K, powder X-ray diffraction figures, crystal structure figures and DFT calculations.

**Accession Codes:**

CCDC 2144723 contain the supplementary crystallographic data for this paper. These data can be obtained free of charge via www.ccdc.cam.ac.uk/data_request/cif, or by emailing data_request@ccdc.cam.ac.uk, or by contacting The Cambridge Crystallographic Data Centre, 12 Union Road, Cambridge CB2 1EZ, UK; fax: +44 1223 336033.

# AUTHOR INFORMATION

**Corresponding Author**

Liurukara D. Sanjeewa (sanjeewal@missouri.edu)

**Acknowledgements**:

This research used resources at the Missouri University Research Reactor (MURR). This work was supported in part by a University of Missouri Research Council Grant (Grant Number: URC-22-021). The research at the Oak Ridge National Laboratory (ORNL) is supported by the U.S. Department of Energy (DOE), Office of Science, Basic Energy Sciences (BES), Materials Sciences and Engineering Division. This research used resources at the High Flux Isotope Reactor and Spallation Neutron Source, DOE Office of Science User Facilities operated by ORNL.